\documentclass{kluwer}    

\newdisplay{guess}{Conjecture}

\begin{document}
\begin{article}
\begin{opening}
\title{Formation of Small-Scale Condensations in the Molecular Clouds via Thermal
Instability}

\author{Mohsen \surname{Nejad-Asghar}$^1$
\footnote{nasghar@dubs.ac.ir} \& Jamshid \surname{Ghanbari}$^2$}

\runningauthor{Nejad-Asghar \& Ghanbari}

\runningtitle{Small-Scale Condensation Formation}

\institute{$^1$School of Physics, Damghan University of Basic
            Sciences, Iran \\ $^2$Department of Physics, School of
Sciences,Ferdowsi University, Mashhad, Iran}

\date{received ... accepted ..., in original form ...}

\begin{abstract}
A systematic study of the linear thermal instability of a
self-gravitating magnetic molecular cloud is carried out for the
case when the unperturbed background is subject to local
expansion or contraction. We consider the ambipolar diffusion, or
ion-neutral friction on the perturbed states. In this way, we
obtain a non-dimensional characteristic equation that reduces to
the prior characteristic equation in the non-gravitating
stationary background. By parametric manipulation of this
characteristic equation, we conclude that there are, not only
oblate condensation forming solutions, but also prolate solutions
according to local expansion or contraction of the background. We
obtain the conditions for existence of the Field lengths that
thermal instability in the molecular clouds can occur. If these
conditions establish, small-scale condensations in the form of
spherical, oblate, or prolate may be produced via thermal
instability.
\end{abstract}

\end{opening}

\section{Introduction}

With the increase in the observational resolution, smaller sizes
of density fluctuations have been detected. Direct imaging of
$^{12}CO$ in nearby clouds revealed substructures with scales of
$\sim0.01pc$ ($\sim0.01M_{\odot}$) (Peng et al. 1998, Sakamoto \&
Sunada 2003). Studies of the time variability of absorption lines
indicates the presence of small-scale clumps in the dense gas on
scales so small as $\sim5\times10^{-5}pc$ ($\sim10AU$), with
masses of $\sim5\times10^{-9}M_{\odot}$ (Moore \& Marscher 1995,
Rollinde et al. 2003).

As a general rule, neither subparsec nor AU-scale condensations
in molecular clouds are spherical (Ryden 1996). Jones \& Basu
(2002) have recently decipher intrinsic three-dimensional shape
distributions of molecular clouds, cloud cores, Bok globules, and
condensations. They find out that molecular clouds mapped in
$^{12}CO$ are intrinsically triaxial but more nearly prolate than
oblate, while the smaller cloud cores, Bok globules, and
small-scale condensations are also intrinsically triaxial but
more nearly oblate than prolate.

Small-scale condensations appear to be immediate precursor of
large-scale clumps (dense cores with significant jeans mass) via
merging and collisions; they constitute the initial conditions
for star formation. Therefore, the understanding of the origin
and merging of these small-scale condensations is of fundamental
importance for a consistent theory of star formation and galactic
evolution.

The origin and shape of these small-scale condensations is a
disputable issue. Anisotropic heating and fragmentation via
gravitational collapse is an important reason for oblate/prolate
large-scale clumps with significant Jeans mass (e.g. Nelson \&
Langer 1997, Indebetouw \& Zweibel 2000, Hartmann 2002). The
above scenario is not correct for small-scale condensations,
because they have low gas density and small sizes, thus, their
masses are significantly smaller than their corresponding Jeans
mass. According to this feature, the only remaining responsible
parameters may be \textit{turbulence} and/or \textit{thermal
instability}.

Gammie et al. (2003) have recently studied the effect of
turbulence in three dimensional analogs of clumps using a set of
self consistent, time-dependent, numerical models of molecular
clouds. The models follow the decay of initially supersonic
turbulence in an isothermal, self-gravitating, magnetized fluid.
They have concluded that nearly $90\%$ of the clumps are formed in
prolate and $10\%$ of them are oblate.

In molecular clouds, the dispersion velocity inferred from
molecular line width is often larger than the gas sound speed
inferred from transition temperatures (Solomon et al 1987).
Magnetohydrodynamic turbulence may be responsible for the
stirring of these clouds (Arons \& Max 1975). Because of these
turbulent motions, molecular clouds must be transient structures,
and are probably dispersed after not much more than $\sim 10^7 yr$
(Larson 1981). Since cooling time-scale of molecular clouds is
approximately $\sim 10^3-10^4 yr$ (Gilden 1984), thermal
instability may be a coordinated trigger mechanism to form
condensations. Turbulence, in the second stage, can deform these
small-scale condensations in shape and orient them relative to
the background magnetic fields.

Observations and theoretical studies establish that magnetic
fields play an important role in shaping the structure and
dynamics of molecular clouds and their substructures (e.g. Basu
2000, Fiege \& Pudritz 2000, Hennebelle 2003). The relative
alignment of the projected magnetic field with the projected
minor axis of the condensations is an important diagnostic.

In conformity with the above explanation, Nejad-Asghar \& Ghanbari
(2003 hereafter NG) interested to investigate the effect of
ambipolar diffusion on the thermal instability and formation of
small-scale condensations in the magnetic molecular clouds. They
concluded that there are solutions where the thermal instability
allows compression along the magnetic field but not perpendicular
to it. NG inferred that this aspect might be evidence in
formation of the observed oblate small-scale condensations in
magnetic molecular clouds.

In this paper we want to testify and develop the work of NG by
including self-gravity and local background contraction/expansion.
We present the basic equations, background evolution, and the
linearized equations in section 2. Section 3 deals with
exponential growth rate and parametric solutions that culminates
in formation of oblate, prolate, and spherical condensations.
Section 4 allocates to a conclusion and some future prospects.

\section{The Equations of The Problem}

The basic equations, including self-gravity and ambipolar
diffusion, are given first in general (\S~2.1) and then
specialized for the homogeneous contracting/expanding molecular
cloud (\S~2.2) and for small perturbations to that medium
(\S~2.3).
\subsection{Equations}

In principle, the ion velocity $\textbf{\textit{v}}_i$ and the
neutral velocity $\textbf{\textit{v}}_n$ in molecular clouds,
should be determined by solving separate fluid equations for
these species (Draine 1986), including their coupling by collision
processes. But, in the time-scale of cooling considered here,
($10^3-10^4 yr$, Gilden 1984), two fluids of ion and neutral are
well coupled together, and we can use the basic equations as
follows (Shu 1992)
\begin{equation}\label{e:basic1}
\frac{d\rho}{dt} + \rho\nabla\cdot\textbf{\textit{v}}=0
\end{equation}
\begin{equation}
\rho\frac{d\textbf{\textit{v}}}{dt} + \nabla p +
\nabla(\frac{B^2}{8\pi}) -
(\textbf{B}\cdot\nabla)\frac{\textbf{B}}{4\pi}+\rho\nabla\psi=0
\end{equation}
\begin{equation}
\frac{1}{\gamma-1}\frac{dp}{dt} -
\frac{\gamma}{\gamma-1}\frac{p}{\rho}\frac{d{\rho}}{dt} +
\rho\Omega-\nabla\cdot(K\nabla T)=0
\end{equation}
\begin{equation}
\frac{d\textbf{B}}{dt}+\textbf{B}(\nabla\cdot\textbf{\textit{v}})-
(\textbf{B}\cdot\nabla)\textbf{\textit{v}}=
\nabla\times\{\frac{\textbf{B}}{4\pi\eta\epsilon\rho^{1+\nu}}\times
[\textbf{B}\times(\nabla\times\textbf{B})]\}
\end{equation}
\begin{equation}
\nabla^2\psi=4\pi G\rho
\end{equation}
\begin{equation}\label{e:basic6}
p-\frac{R}{\mu}\rho T=0
\end{equation}
where variables and parameters have their usual meanings and
$\eta\approx 2.46\times 10^{14}cm^3.g^{-1}.s^{-1}$ is the
collision drag in molecular clouds (see McDaniel \& Mason 1973).
We use the relation $\rho_i=\epsilon\rho_n^\nu$
($\epsilon\approx1.83\times10^{-17} cm^{-3/2}.g^{1/2}$,
$\nu=1/2$) between ion and neutral densities (Umebayasi \& Nakano
1980), and in a good approximation we choose
$\rho=\rho_n+\rho_i\approx\rho_n$.

$\Omega(\rho,T)=\Lambda(\rho,T)-\Gamma_{tot}$ is the net cooling
function ($erg.s^{-1}.g^{-1}$), where $\Gamma_{tot}$ is the total
heating rate and $\Lambda(\rho,T)$ is the cooling rate which can
be written as (Goldsmith \& Langer 1978, Neufeld et al. 1995)
\begin{equation}
\Lambda(\rho,T)=\Lambda_0 \rho^\delta T^\beta
\end{equation}
where $\Lambda_0$, $\delta$, and $\beta$ are constants. The range
of $\beta$ is $1.4$ to $2.9$. The constant $\delta$ is greater
than zero for optically thin case and less than zero for
optically thick case (see Fig.~1).

\input{epsf}
\begin{figure} 
 \centerline{{\epsfxsize=3in\epsfysize=4in\epsffile{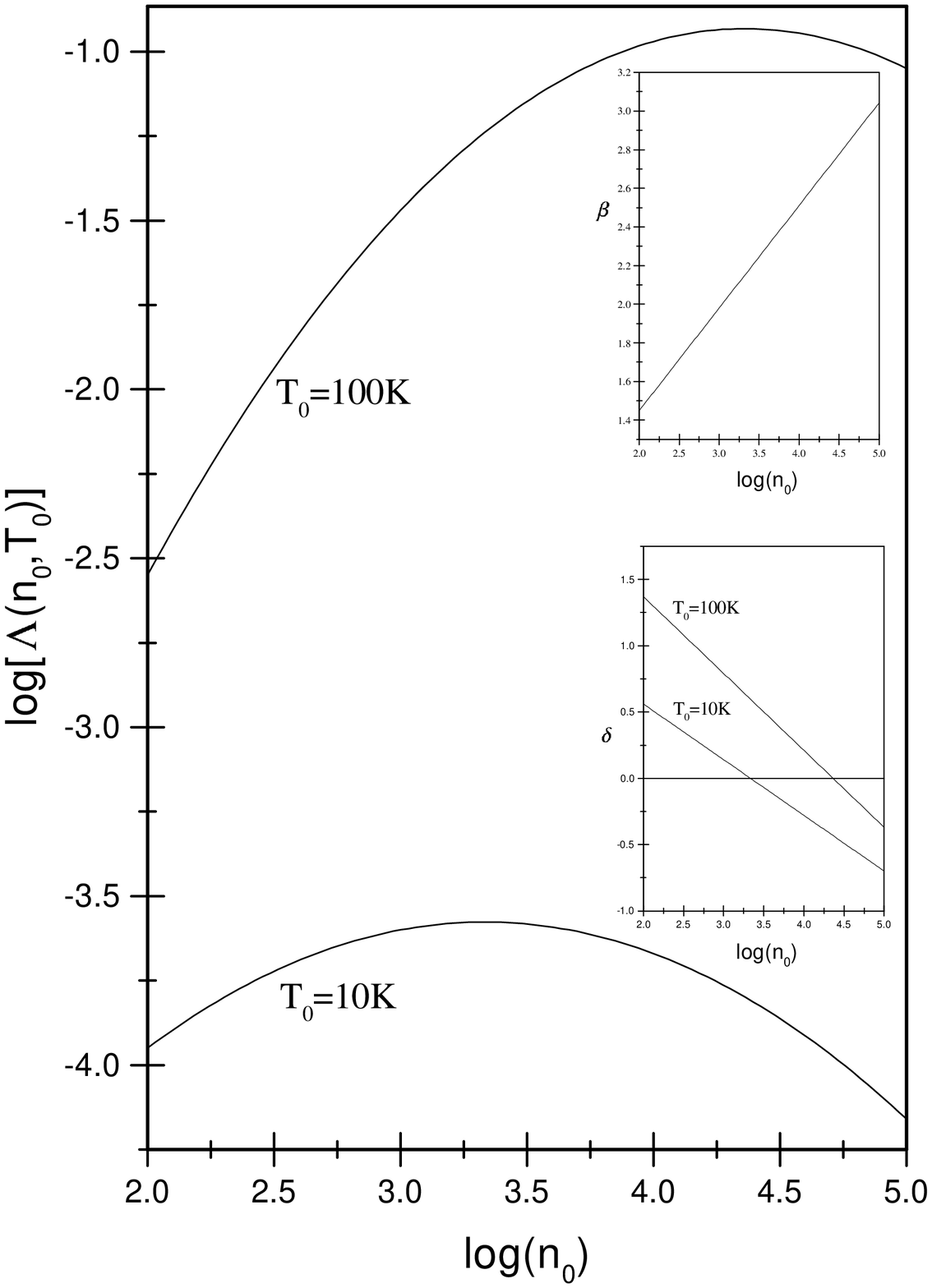}}}
 \caption[]{Logarithm of the cooling rate, $\Lambda(n_0,T_0)=
  \Lambda_0n_0^\delta T_0^\beta$, versus number density in molecular
  clouds, $n_0(cm^{-3})$. The values of $\delta$ and $\beta$ are shown
  in this figure.}
\end{figure}

Models of the molecular clouds identify several different heating
mechanisms. In this paper, we consider the heating rates of
cosmic rays, $H_2$ formation, $H_2$ dissociation, grain
photoelectrons, and collisions with warm dust, as a constant
$\Gamma_0$ (Glassgold \& Langer 1974, Goldsmith \& Langer 1978).
The heating of the gas by magnetic ion-neutral slip is discussed
in detail by Scalo(1977); a simple estimate of this heating rate
is
\begin{equation}
\Gamma_{AD}=\eta \epsilon \rho^\nu \textit{v}_d^2
\end{equation}
where $\textit{v}_d$ is the drift velocity of ions
\begin{equation}
\textit{v}_d=\frac{1}{4\pi\eta\epsilon\rho^{1+\nu}}
\mid(\nabla\times\textbf{B})\times\textbf{B}\mid.
\end{equation}
In order of magnitude, if $\kappa B_0$ changes on a typical scale
of $\lambda$, then
\begin{equation}
\textit{v}_d\sim\frac{(\kappa
B_0)^2}{4\pi\eta\epsilon\rho^{1+\nu}\lambda},
\end{equation}
and ambipolar diffusion heating rate is given by
\begin{equation}
\Gamma_{AD}=\Gamma_0'\rho^{-(2+\nu)}
\end{equation}
where $\Gamma_0'$ is defined as
\begin{equation}
\Gamma_0'\equiv\frac{(\kappa
B_0)^4}{16\pi^2\eta\epsilon\lambda^2}.
\end{equation}
The gravitational heating rate is found by setting the rate of
compressional/expansional work per particle, $pd(n^{-1})/dt$,
equal to the rate of change of gravitational energy per particle,
$[d(PE_{tot})/dt]/(nV)$, where $PE_{tot}$ is the gravitational
potential energy of the volume $V$. For a uniform sphere of radius
$\lambda$, we find
\begin{equation}
\Gamma_{grav}=\Gamma_0''\rho^{3/2}
\end{equation}
where $\Gamma_0''$ is defined as
\begin{equation}
\Gamma_0''\equiv\frac{(4\pi
G)^{3/2}}{5\sqrt{3}}[-\dot{a}(\tau)]\lambda^2
\end{equation}
where $\dot{a}(\tau)$ is the contraction/expansion parameter rate
(see \S 2.2).

\subsection{Background Evolution}

As a basis for the small-perturbation analysis, we assume a
homogeneous background which is expanding/contracting uniformly,
so that the unperturbed quantities only depend on time. The
background quantities will be denoted with the subscript $0$. The
expansion is given by
\begin{equation}\label{e:lag}
\textbf{\textit{r}}=a(t)\textbf{\textit{x}}
\end{equation}
where $\textbf{\textit{r}}$ is the Eulerian coordinate,
$\textbf{\textit{x}}$ is the Lagrangian coordinate and $a(t)$ is
the expansion/contraction parameter. Using equation (\ref{e:lag}),
the unperturbed velocity field is given by
\begin{equation}
\textbf{\textit{v}}_0(\textbf{\textit{r}},t)=\frac{da/dt}{a}\textbf{\textit{r}}.
\end{equation}
for the background evolution, the basic equations
(\ref{e:basic1})-(\ref{e:basic6}) reduce to
\begin{eqnarray}\label{e:back}
\nonumber{\rho_0(t)=\rho_0(t=0)a(t)^{-3},\quad
p_0(t)=p_0(t=0)a(t)^{-3\gamma}}\\
T_0(t)=T_0(t=0)a(t)^{-3(\gamma-1)},\quad
\textbf{B}_0(t)=\textbf{B}_0(t=0)a(t)^{-2}
\end{eqnarray}
where $a(t)$ follows the differential equation
\begin{equation}
a(\tau)^2\ddot{a}(\tau)=-1
\end{equation}
where the dot over the symbol indicates the derivative respect to
a non-dimensional variable $\tau\equiv[\frac{4}{3}\pi
G\rho_0(t=0)]^{1/2}t$.
\subsection{Linearized Equations}

Density fluctuation ratios in the molecular substructures is in
the order of $\sim 10$ (Falgarone et al. 1992, Pan et al. 2001).
Therefore, the linear regime of the thermal instability might lead
to some significant results for small-scale condensation
formation.

To obtain a linearized system of equations, we split each
variable into unperturbed and perturbed components, indicating
the latter with a subscript $1$. Eulerian del operator is applied
to the equations and all equations are rewritten in terms of the
Lagrangian coordinate $\textbf{\textit{x}}$. The resulting linear
system has coefficients which depend on $t$ but not on
$\textbf{\textit{x}}$. We then carry out a spatial Fourier
analysis, with Fourier components proportional to
$exp(i\textbf{\textit{k}}\cdot\textbf{\textit{x}})$, so that
$\textbf{\textit{k}}$ is the Lagrangian wave vector.

Simplifying the resulted linear system by repeated use of the
background equations (\ref{e:back}); we obtain
\begin{equation}
\frac{d}{dt}(\frac{\rho_1}{\rho_0})+i\tilde{\textbf{\textit{k}}}\cdot
\textbf{\textit{v}}_1=0
\end{equation}
\begin{eqnarray}\label{e:linear1}
\nonumber{\frac{d\textbf{\textit{v}}_1}{dt}+\frac{da/at}{a}\textbf{\textit{v}}_1
+i\frac{c_s}{\gamma\tau_s}\frac{\tilde{\textbf{\textit{k}}}}{\tilde{\textit{k}}}
(\frac{p_1}{p_0})+i\frac{\textbf{B}_0\cdot\textbf{B}_1}{4\pi\rho_0}
\tilde{\textbf{\textit{k}}}} \\
-i\frac{\tilde{\textbf{\textit{k}}}\cdot\textbf{B}_0}
{4\pi\rho_0}\textbf{B}_1-i\frac{\tilde{\textbf{\textit{k}}}}
{\tilde{\textit{k}}^2\tau_g^2}(\frac{\rho_1}{\rho_0})\lefteqn{=0}
\end{eqnarray}
\begin{equation}
\frac{d}{dt}(\frac{p_1}{p_0})+i\gamma\tilde{\textbf{\textit{k}}}
\cdot\textbf{\textit{v}}_1+(\frac{1}{\tau_{c\rho}}+\frac{1}{\tau_K})(\frac{p_1}{p_0})
+(\frac{1}{\tau_{cT}}-\frac{1}{\tau_{c\rho}}-\frac{1}{\tau_K})(\frac{\rho_1}{\rho_0})=0
\end{equation}
\begin{eqnarray}\label{e:linear2}
\nonumber{\frac{d\textbf{B}_1}{dt}
+i\textbf{B}_0(\tilde{\textbf{\textit{k}}}
\cdot\textbf{\textit{v}}_1)-
i(\tilde{\textbf{\textit{k}}}\cdot\textbf{B}_0)\textbf{\textit{v}}_1
+\frac{2da/dt}{a}\textbf{B}_1} \\
+\tilde{\textbf{\textit{k}}}\times\{\frac{\textbf{B}_0}{4\pi\eta\epsilon\rho_0^{1+\nu}}
\times[\textbf{B}_0\times(\tilde{\textbf{\textit{k}}}\times\textbf{B}_1)]\}\lefteqn{=0}
\end{eqnarray}
where $c_s=\sqrt{\gamma p_0/\rho_0}$ and
$\tilde{\textbf{\textit{k}}}=\textbf{\textit{k}}/a$ are,
respectively, the adiabatic background sound speed and the
Eulerian wave vector. The other symbols have the following
definitions:
\begin{eqnarray}\label{e:time}
\nonumber{\tau_s\equiv\frac{1}{\tilde{\textit{k}}c_s},
\quad\tau_g\equiv\frac{1}{\sqrt{4\pi G\rho_0}},
\quad\tau_K\equiv\frac{R\rho_0}{\mu(\gamma-1)K\tilde{\textit{k}}^2},}\\
\tau_{cT}\equiv\frac{RT_0}{\mu(\gamma-1)
\rho_0(\partial\Omega/\partial\rho)_T},
\quad\tau_{c\rho}\equiv\frac{R}{\mu(\gamma-1)
(\partial\Omega/\partial T)_\rho};
\end{eqnarray}
that are the characteristic time-scale of sound waves,
self-gravity perturbation waves, thermal conduction, isothermal
and isobaric differential cooling, respectively.

We use the coordinate system $\textbf{\textit{u}}_x$,
$\textbf{\textit{u}}_y$, and $\textbf{\textit{u}}_z$ as specified
by NG. Equations (\ref{e:linear1}) and (\ref{e:linear2}) may be
used to uncouple $\textit{v}_{1y}$- the perturbed velocity in the
plane perpendicular to both $\textbf{B}_0$ and
$\textbf{\textit{k}}$- from the rest of the problem. With the
choice of exponential perturbation ($e^{ht}$), disturbances
perpendicular to the $(\textbf{B}_0-\textbf{\textit{k}})$-plane,
have a solution which displays existence or non-existence of the
Alfv\'{e}n waves. Amplitude of the Alfv\'{e}n waves are damped
via expansion of the medium and/or with ion-neutral friction,
while, it must grow with injection of energy in contracting
medium.

The motion in the other modes are constrained to the $xz$-plane,
and are governed by the matrix equation,
\begin{equation}\label{e:matrix}
Y^{(1)}=AY
\end{equation}
where $Y$ is a $5\times 1$ matrix as
\[ Y= \left(
        \begin{array}{c}
             \rho_1 / \rho_0  \\
             p_1 / p_0  \\
             a\textit{v}_{1x}  \\
             a\textit{v}_{1z}  \\
             \sin\theta(\frac{B_{1z}}{B_0})-\cos\theta(\frac{B_{1x}}{B_0})
         \end{array} \right), \]
and $Y^{(1)}$ is its first time derivative. The $5\times 5$
matrix of the coefficients, $A$, is defined as
\[ A= \left(
        \begin{array}{ccccc}
             0 & 0 & -\frac{i\sin\theta}{c_s\tau_sa} & -\frac{i\cos\theta}{c_s\tau_sa} & 0\\
             \frac{1}{\tau_{c\rho}}+\frac{1}{\tau_{K}}-\frac{1}{\tau_{cT}} &
             -\frac{1}{\tau_{c\rho}}-\frac{1}{\tau_{K}} &
             -\frac{i\gamma\sin\theta}{c_s\tau_sa} & -\frac{i\gamma\cos\theta}{c_s\tau_sa} & 0\\
             \frac{ic_s\tau_sa\sin\theta}{\tau_g^2} & \frac{ic_sa\sin\theta}{\gamma\tau_s} &
             0 & 0 & -\frac{ic_s\tau_sa}{\tau_{AL}^2}\\
             \frac{ic_s\tau_sa\cos\theta}{\tau_g^2} & \frac{ic_sa\cos\theta}{\gamma\tau_s} &
             0 & 0 & 0\\
             0 & 0 & -\frac{i}{c_s\tau_sa} & 0 & -\frac{1}{\kappa^2\tau_{AD}}
         \end{array} \right), \]
where $\theta$ is the angle between $\textbf{\textit{k}}$ and
$\textbf{B}_0$, and
\begin{equation}
\tau_{AL}\equiv\frac{1}{\tilde{\textit{k}}\textit{v}_A}=
\frac{\sqrt{4\pi\rho_0}}{\tilde{\textit{k}}B_0},\quad
\tau_{AD}\equiv\frac{1}{\tilde{\textit{k}}\textit{v}_d}=
\frac{4\pi\eta\epsilon\rho_0^{1+\nu}}{\tilde{\textit{k}}^2(\kappa
B_0)^2}
\end{equation}
are the characteristic time-scales of the Alfv\'{e}n waves and
ambipolar diffusion, respectively.
\section{Exponential Growth Rate}

The standard exponential growth rate provides the following
formal solution for all the perturbations:
\begin{equation}\label{e:exp}
y_i(t)=y_i(t=0)exp(ht),
\end{equation}
where $real(h)$ represents the growth/decay rate. Inserting the
background evolution (eq.~[\ref{e:back}]) and exponential growth
form (eq.~[\ref{e:exp}]), into equation (\ref{e:matrix}); we
obtain
\begin{equation}\label{e:linmax}
Y^{(1)}=hY+CY
\end{equation}
where $C$ is a diagonal matrix as
\begin{equation}
C\equiv\frac{1}{\tau_e}diag[3,3\gamma,1,1,2]
\end{equation}
where $\mid \tau_e \mid=\frac{a}{\mid da/dt \mid}$ represent
contraction/expansion time-scale. Existence of solution for
equation (\ref{e:linmax}), needs the following condition:
\begin{equation}
Det[hI+C-A]=0
\end{equation}
where $I$ is the unitary matrix. By introducing the
non-dimensional quantities
\begin{eqnarray}
\nonumber{y\equiv h\tau_s,\;
\sigma_\rho\equiv\frac{\tau_s}{\tau_{cT}},\;
\sigma_T\equiv\frac{\tau_s}{\tau_{c\rho}}
+\frac{\tau_s}{\tau_K},\;}\\
\alpha\equiv(\frac{\tau_s}{\tau_{AL}})^2,\;
D\equiv\frac{\tau_s}{\kappa^2\tau_{AD}},\;
G_g\equiv(\frac{\tau_s}{\tau_g})^2,\;
E_e\equiv\frac{\tau_s}{\tau_e},
\end{eqnarray}
we find a five-degree linear characteristic equation that without
self-gravity and expansion/contraction of the background ($G_g=0$,
$E_e=0$), reduces to the equation (22) of NG. We use the Laguerre
method to find the roots of this characteristic equation.

\subsection{The Sound Domain}

When the sound period, $\tau_s$, is much smaller than the other
characteristic time-scales, we can neglect the effect of
$\alpha$, $D$, $G_g$, and $E_e$. The characteristic equation
reduces to a three-degree linear equation. There are three
solutions: two sound waves and one condensation mode. The stable
region for this case is shown in the Fig.~1 of NG.

For the temperatures and densities that thermal instability is
destabilizing, there exist a critical length (Field length)
defined as the maximum wavelength which thermal conduction can
suppress the instability. Inserting the defined time-scales
(eq.~[\ref{e:time}]), into the definitions of $\sigma_\rho$ and
$\sigma_T$; we obtain
\begin{equation}
\sigma_\rho=\frac{\mu(\gamma-1)}
{RT_0\tilde{k}c_s}\Lambda(n_0,T_0)(\delta+2.5\xi-1.5\chi)
\end{equation}
\begin{equation}
\sigma_T=\frac{\mu(\gamma-1)}
{RT_0\tilde{k}c_s}\Lambda(n_0,T_0)\beta
[1+(\frac{\lambda_0}{\lambda})^2]
\end{equation}
where $\xi$ and $\chi$ are
\begin{equation}
\xi\equiv\frac{\Gamma_{AD}}{\Lambda(n_0,T_0)},\quad
\chi\equiv\frac{\Gamma_{grav}}{\Lambda(n_0,T_0)}
\end{equation}
and $\lambda_0$ is a defined wavelength as follows:
\begin{equation}\label{e:lam0}
\lambda_0\equiv\sqrt{\frac{KT_0}{\beta n_0\Lambda(n_0,T_0)}}.
\end{equation}
The values of the $\lambda_{0(pc)}$ for typical data in the
molecular clouds is given in Fig.~2. According to this figure, we
choose wavelengths in the range of $\lambda_{(pc)}\approx
10^{-4}-10^{-1}pc$, that are interesting in formation of the
small-scale condensations.

\input{epsf}
\begin{figure} 
 \centerline{{\epsfxsize=2in\epsfysize=2.5in\epsffile{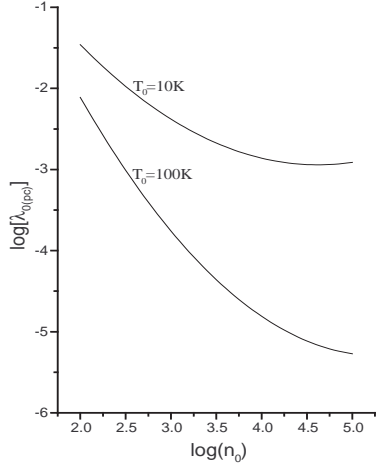}}}
 \caption[]{Logarithm of the Field length (at unit of parsec) in the
 typical molecular clouds versus number density, $n_0(cm^{-3})$.}
\end{figure}

We can now define a generalized Field length for two cases as
follows:
\begin{itemize}
\item $\delta+2.5\xi-1.5\chi>\beta$, which $\sigma_\rho>0$, that is
upwards of the $\sigma_T-\sigma_\rho$ plane:
\begin{equation}\label{e:fi}
\lambda_F^{(i)}=\frac{\lambda_0}
{\sqrt{\frac{\delta+2.5\xi-1.5\chi}{\beta}-1}}
\end{equation}
\item $\delta+2.5\xi-1.5\chi<-\frac{2}{3}\beta$, which $\sigma_\rho<0$, that
is downwards of the $\sigma_T-\sigma_\rho$ plane:
\begin{equation}\label{e:fii}
\lambda_F^{(ii)}=\frac{\lambda_0}
{\sqrt{\frac{3}{2}\frac{1.5\chi-\delta-2.5\xi}{\beta}-1}}.
\end{equation}
\end{itemize}
If $\delta+2.5\xi-1.5\chi$ sets between $-\frac{2}{3}\beta$ and
$\beta$, the Field length can not be defined, thus, the medium is
stable for all wavelengths. Otherwise, for wavelengths greater
than the defined Field lengths (eq.~[\ref{e:fi}] and
[\ref{e:fii}]), the medium is unstable. This case is shown in
Fig.~3 for two temperatures of $10K$ and $100K$.

\input{epsf}
\begin{figure} 
 \centerline{{\epsfxsize=2in\epsfysize=2.8in\epsffile{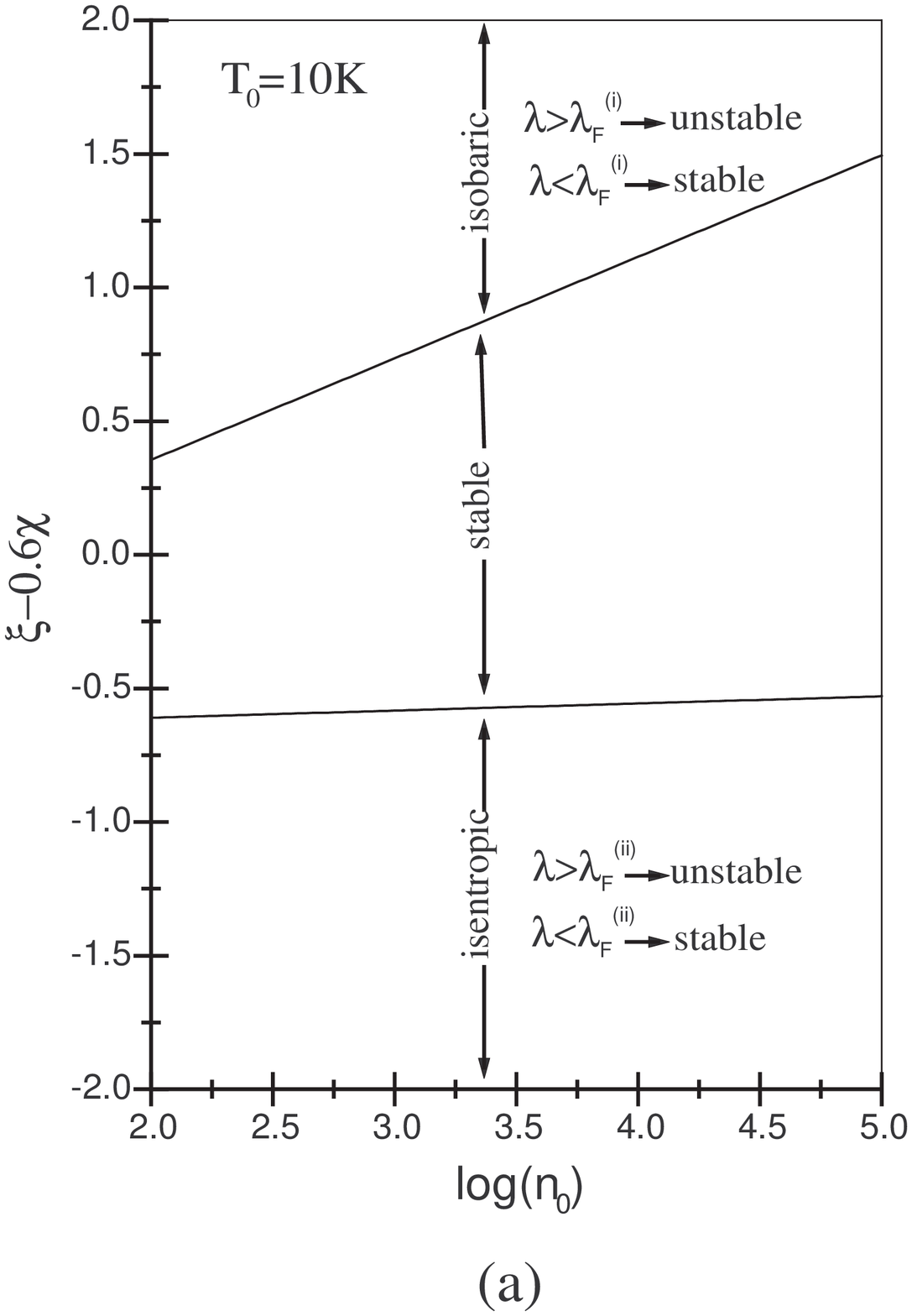}}
   {\epsfxsize=2in\epsfysize=2.8in\epsffile{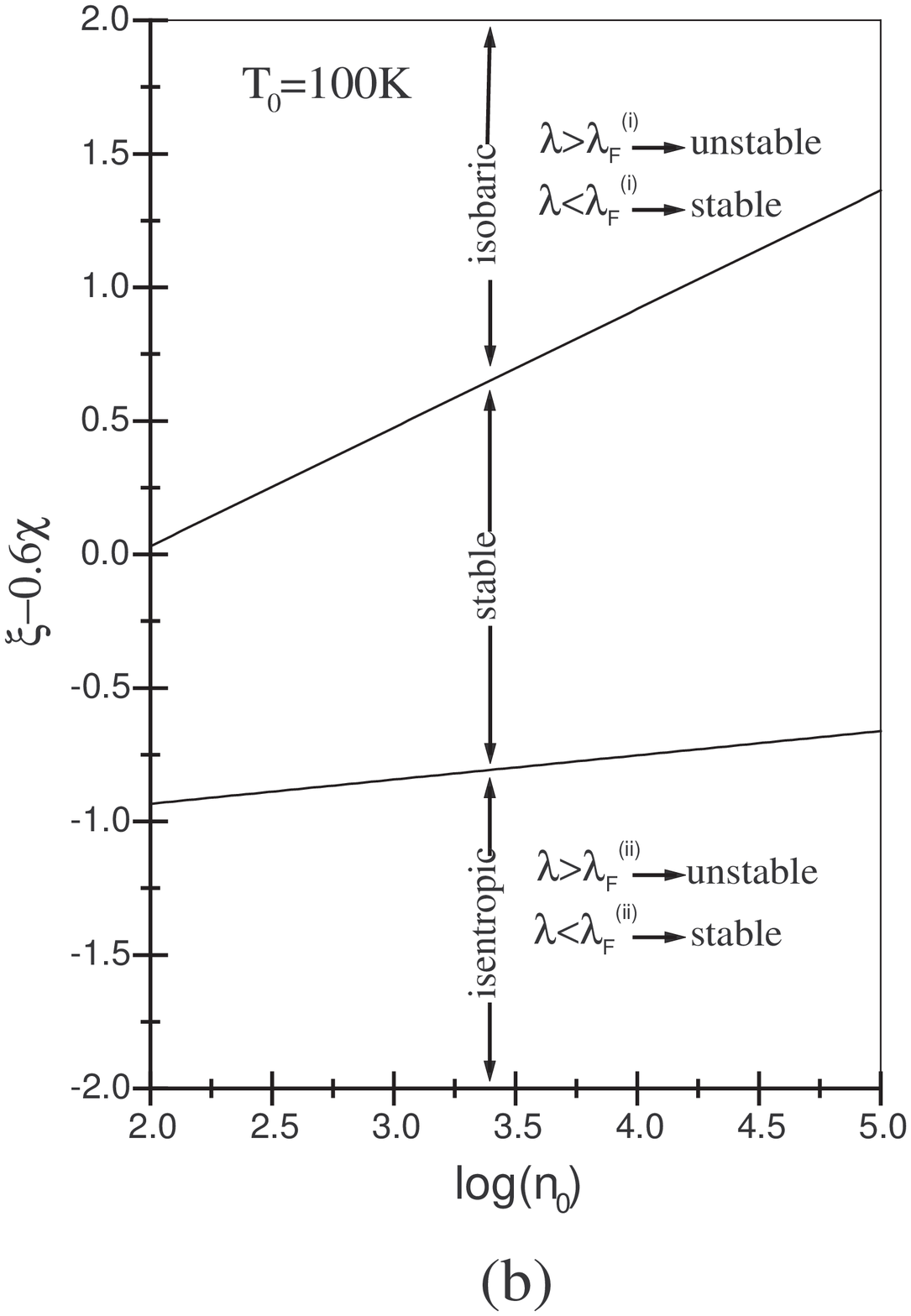}}}
 \caption[]{The stable, isobaric instability, and isentropic instability of
 the medium which can occur according to the different values of
 $\xi-0.6\chi$, for (a) $T_0=10K$ and (b) $T_0=100K$.}
\end{figure}

Before proceeding any further, we must have a real physical
feeling of defined non-dimensional parameters: $\alpha$, $D$,
$E_e$, and $G_g$. We consider some typical magnetic molecular
clouds with density between $10^2cm^{-3}$ to $10^5cm^{-3}$,
temperatures in the range of $T_0\approx 10-100K$, and magnetic
field strength $B_0\approx 10 \mu G$ (Myers \& Goodman 1988,
Crutcher 1999). The typical values of $\alpha$, $D$, $E_e$, and
$G_g$ are shown in Fig.~4.

\input{epsf}
\begin{figure} 
 \centerline{{\epsfxsize=3in\epsfysize=4.1in\epsffile{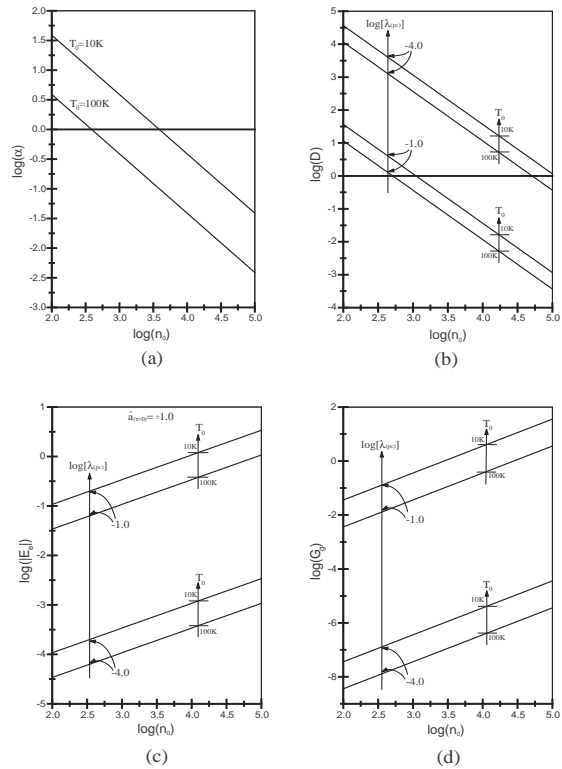}}}
 \caption[]{The typical values of $\alpha$, $D$, $E_e$, and
   $G_g$ in molecular clouds.}
\end{figure}

\subsection{The Magnetic Domain}

Wherever the Alfv\'{e}n period and the ambipolar diffusion
time-scale are important in the magnetic molecular clouds, we can
not ignore the effect of $\alpha$ and $D$ in the characteristic
equation. This case has recently been investigated by NG, without
self-gravity and local expansion/contracion of the background
($G_g\approx0$, $E_e\approx0$). They conclude that there are
solutions where the thermal instability allows compression along
the magnetic field but not perpendicular to it (see Fig.~2 and
Fig.~4 of NG). Maximum effect of the magnetic domain for the
oblate condensation formation is occurred when ambipolar
diffusion time-scale is nearly equal to the period of sound waves
$D\approx 1.0$ (i.e. $\tau_s\approx\kappa^2\tau_{AD}$).

\subsection{The Self-Gravity and Expansion/Contraction Domain}

In this subsection we investigate the effect of the self-gravity
and the expansion/contraction of background. If we consider the
effect of the self gravity without expansion/contraction of the
background, the isobaric  instability criterion (line $OA$ of
Fig.~1 of NG) is modified via bringing this line downwards. This
case is shown in Fig.~5, for a typical value of $\alpha=5.0$ and
$D=1.0$. Physically, this means that self-gravity causes to
increase the internal pressure, thus, isobaric instability must
occur at a decreased $\sigma_\rho$ for each $\sigma_T$.

\input{epsf}
\begin{figure} 
 \centerline{{\epsfxsize=3in\epsfysize=4in\epsffile{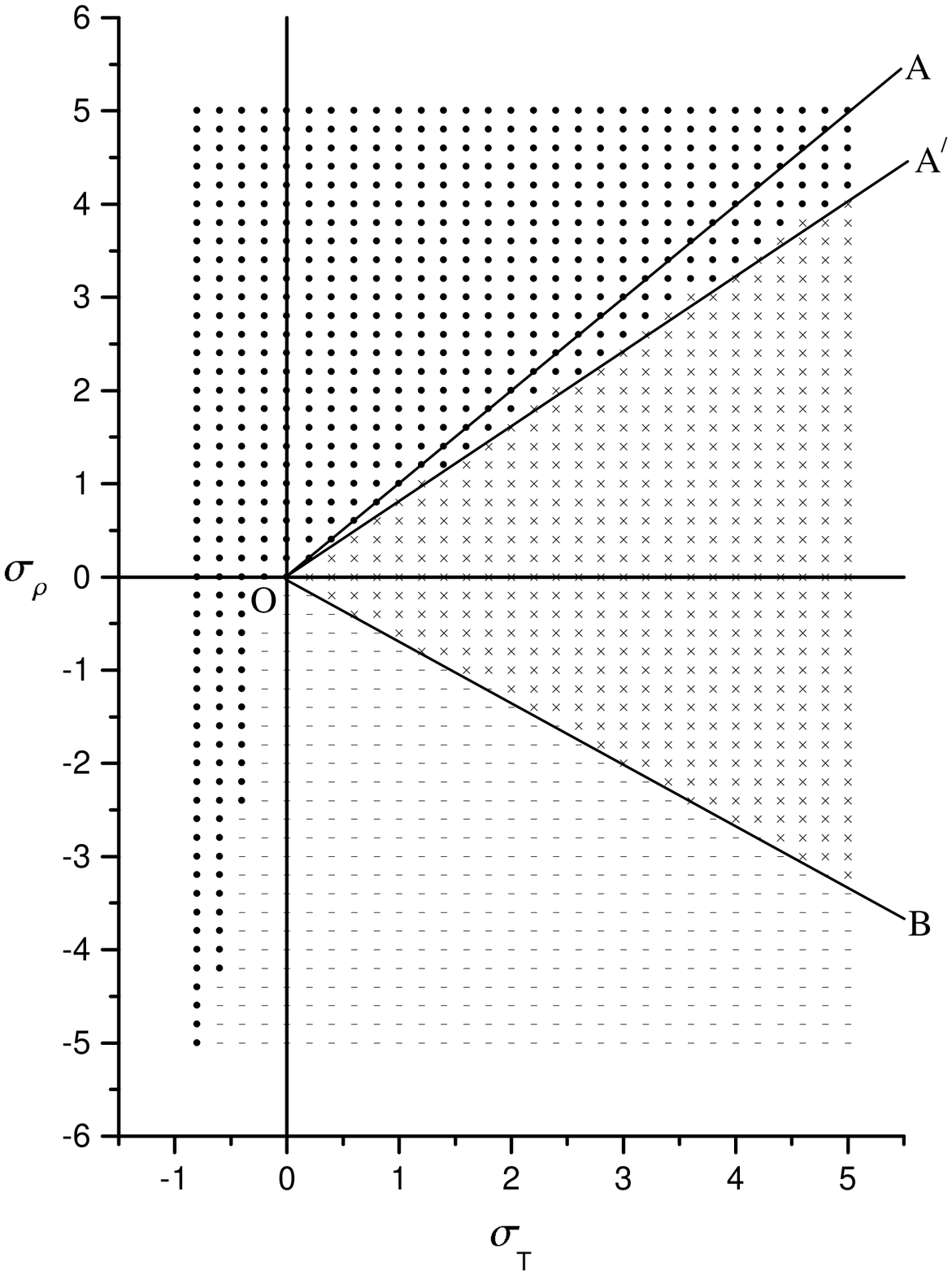}}}
 \caption[]{Regions of stability($\times$),
 spherical instability($\bullet$), and oblate instability($-$)
 for the case of $G_g=0.1$ and $E_e=0$, for a typical value of
 $\alpha=5.0$ and $D=1.0$. The decreased stability region
 in the isobaric instability criterion (line $OA$), occurs by
 the increasing of internal pressure via existence of self-gravity.}
\end{figure}

If the background is expanding ($E_e>0$), its expansion energy
causes to stabilize the medium. This case is shown in Fig.~6 for
a typical value of $\alpha=5.0$ and $D=1.0$. In the isentropic
instability criterion (line $OB$), expansion energy causes to
stabilize the medium in the direction of the magnetic field and
perpendicular to it. On the other hand, in the isobaric
instability criterion (line $OA$), it only causes to stabilize
the medium in perpendicular to the magnetic field, corresponding
to decreased pressure via ion-neutral friction.

For contracting background ($E_e<0$), contraction energy injected
to the medium, thus, its stability is decreased and converted to
a prolate instability. Diffusion of neutrals relative to the
freezed ions in the perpendicular direction of the magnetic field
is the reason of this prolate instability. This case is shown in
Fig.~7 for a typical value of $\alpha=5.0$ and $D=1.0$. When the
parameters of a magnetic molecular cloud set, locally, in this
region of $\sigma_T-\sigma_\rho$ plane, prolate condensation may
be produced via thermal instability.

\input{epsf}
\begin{figure} 
 \centerline{{\epsfxsize=2.5in\epsfysize=3.5in\epsffile{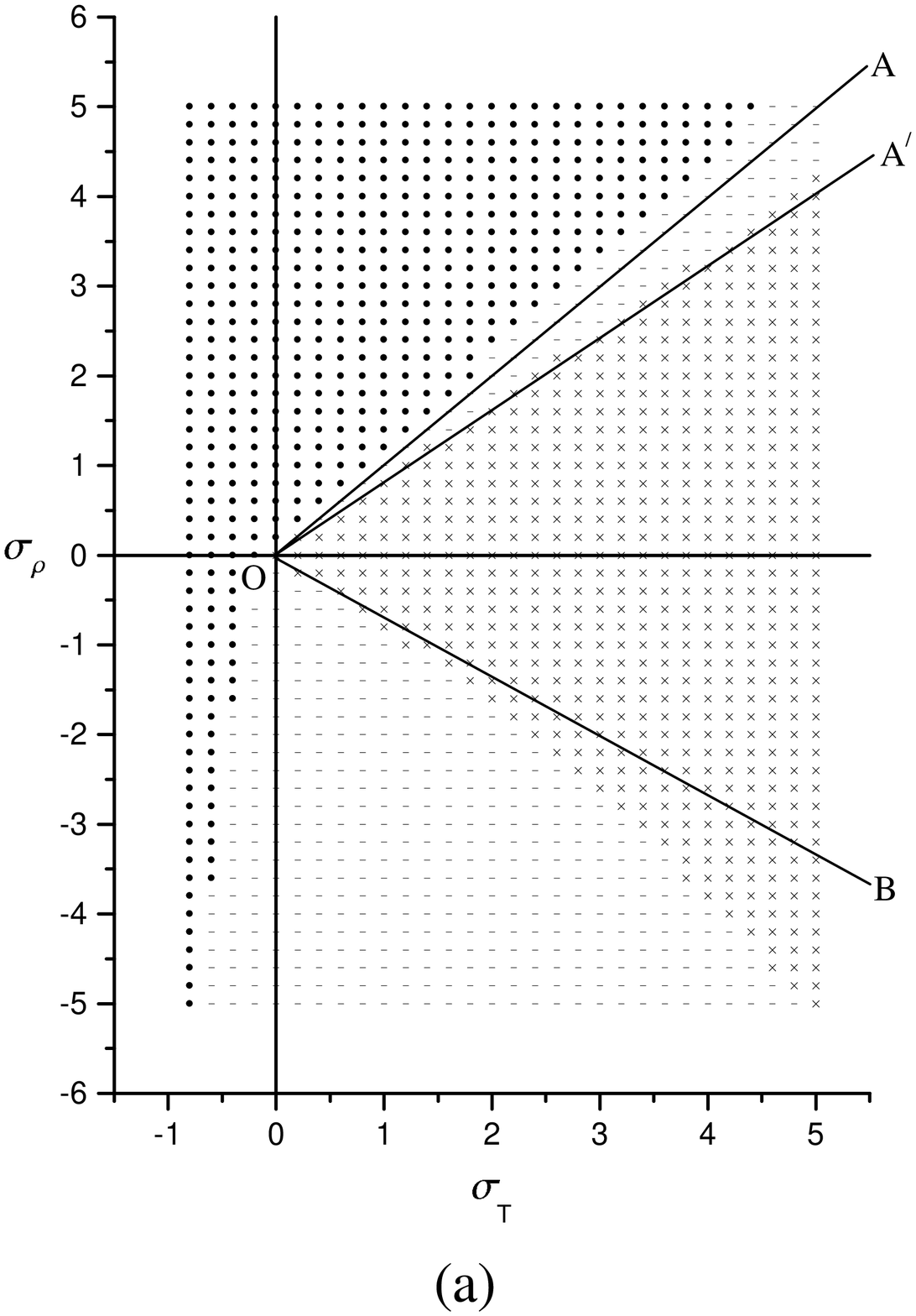}}
   {\epsfxsize=2.5in\epsfysize=3.5in\epsffile{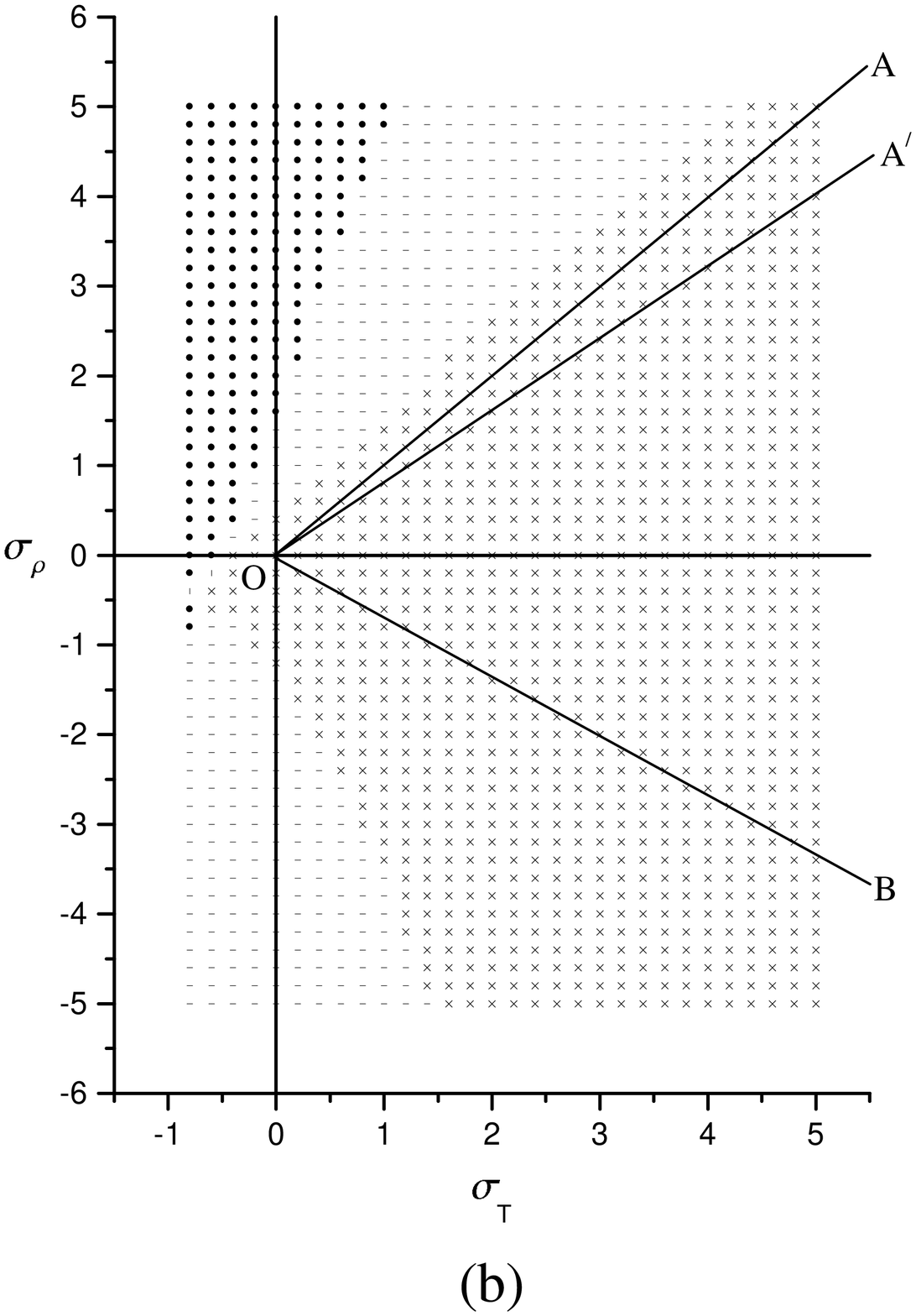}}}
 \caption[]{Regions of stability( $\times$),
 spherical instability($\bullet$), and oblate instability($-$)
 in the expanding background, for a typical value of
 $\alpha=5.0$ and $D=1.0$, with (a) $E_e=0.01$ and $G_g=0.1$, and
 (b) $E_e=0.1$ and $G_g=0$.}
\end{figure}

\begin{figure} 
 \centerline{{\epsfxsize=2.5in\epsfysize=3.5in\epsffile{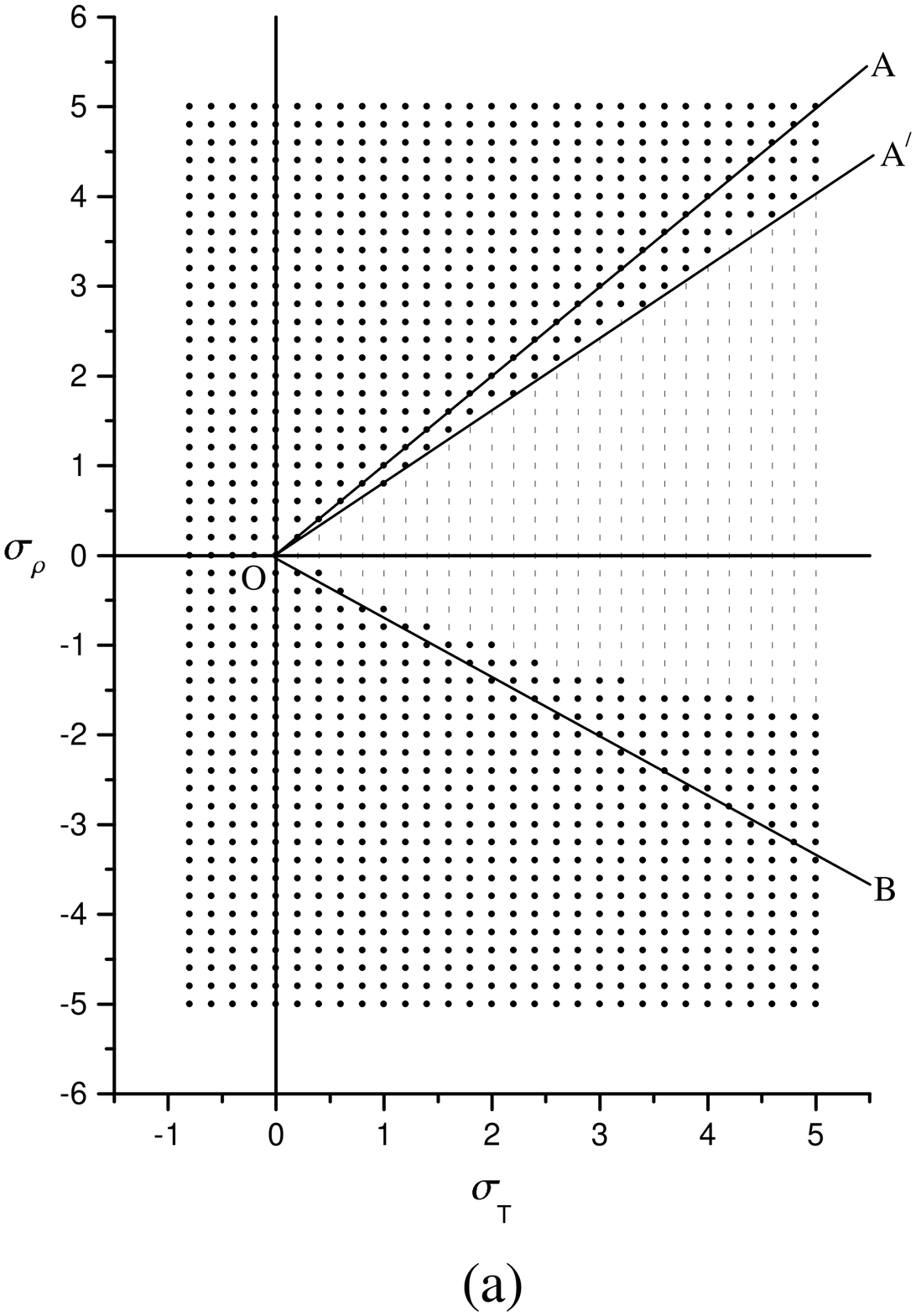}}
   {\epsfxsize=2.5in\epsfysize=3.5in\epsffile{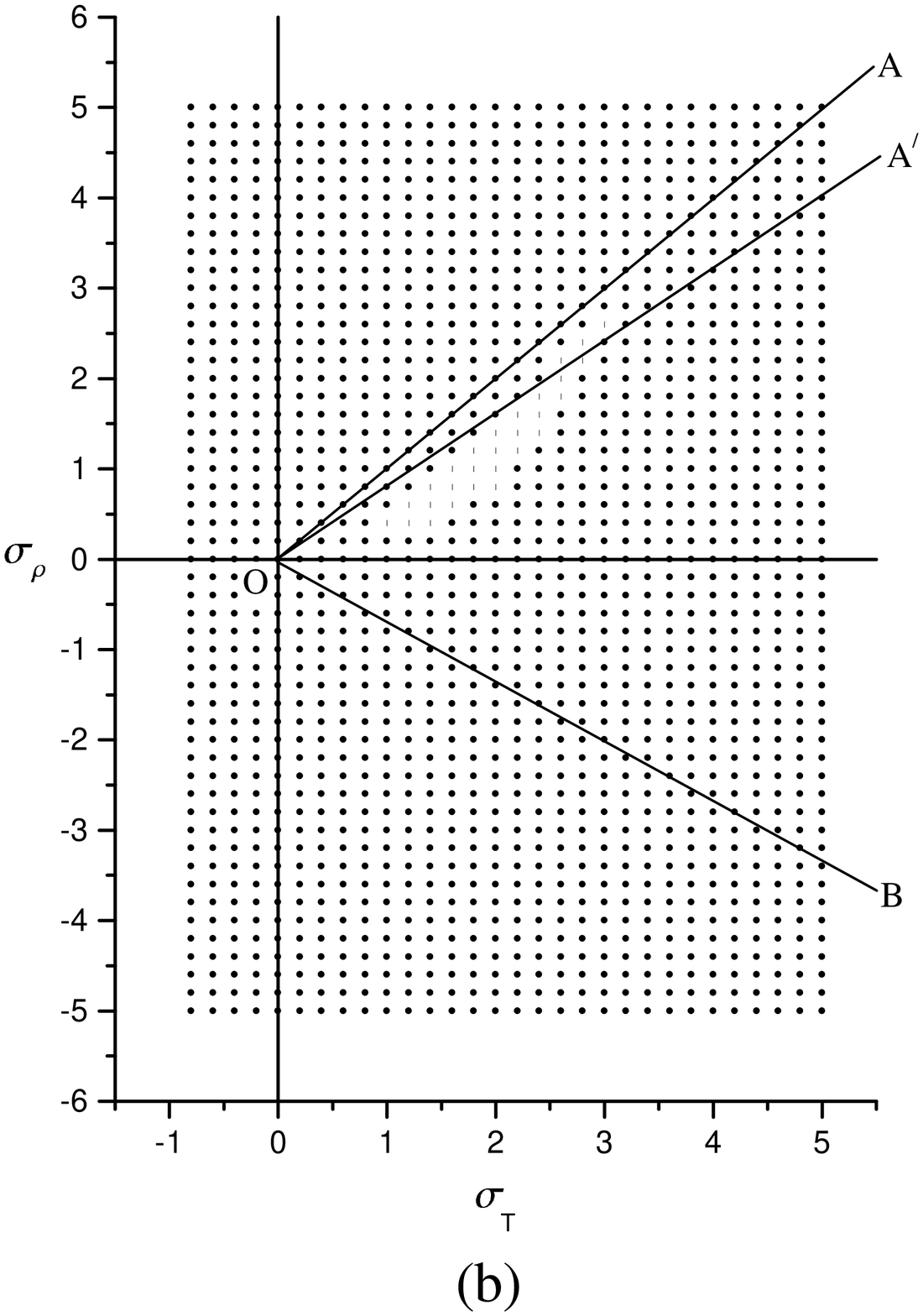}}}
 \caption[]{Regions of spherical instability($\bullet$), and prolate
 instability($|$) in the contracting background, for a typical value of
 $\alpha=5.0$ and $D=1.0$, with (a) $E_e=-0.01$ and $G_g=0.1$, and
 (b) $E_e=-0.1$ and $G_g=0$.}
\end{figure}

\section{Summary And Prospects}

In this paper we perform linear analysis of thermal instability
in a locally uniform expanding/contracting magnetic molecular
clouds which, in the perturbed state, is undergoing ambipolar
diffusion. Thermal conduction and self-gravity have also been
included as fundamental ingredients. The small-perturbation
problem yields a system of ordinary differential equations with
five independent solutions. We choose an exponential growth rate
which convert the system of ordinary differential equations into
a five-degree complete characteristic equation. If we neglect the
self-gravity and expansion/contraction of the background, the
characteristic equation reduces to the prior results of NG. We
have used the Laguerre method to find the roots of this complete
characteristic equation.

In sound domain, two of the solutions have the character of
oscillatory modes (sound waves) and the third one is a
non-oscillatory (or condensation) solution. We adopt a parametric
net cooling function and find for perturbations with wavelengths
greater than the Field length, thermal instability causes the
medium to condense. Fig.~3 shows the condition of instability in
the molecular clouds which their cooling rates are presented in
Fig.~1.

We choose a wide range of density and temperature in the
molecular clouds with typical magnetic field strength $B_0\approx
10 \mu G$. Interesting wavelengths in the problem are around the
Field length which is shown in Fig.~2. According to this figure,
we consider wavelengths around $10^{-4}-10^{-1}pc$ for
small-scale condensations. The typical values of $\alpha$, $D$,
$E_e$, and $G_g$ are shown in Fig.~4.

In magnetic domain, without self-gravity and
expansion/contraction of the background, there are solutions
where the thermal instability allows compression along the
magnetic field but not perpendicular to it. Maximum cases of
these oblate condensation solutions are occurred when the
ambipolar diffusion time-scale equals to the period of sound
waves ($D\approx 1$).

Fig.~5 to Fig.~7 show results that come into existence by
considering of self-gravity and local expansion/contraction of
the magnetic molecular cloud. We deduce that the self-gravity
causes to increase the internal pressure, thus, isobaric
instability must occurs at a decreased $\sigma_\rho$ for each
$\sigma_T$. Therefore, instability of the medium is increased. In
the expanding background, expansion energy in the isentropic
instability, causes to stabilize the medium in the direction of
the magnetic field and perpendicular to it, while, in the
isobaric instability it only stabilize the medium in the
perpendicular direction. In the contracting background, stability
of the medium is decreased and converted to the prolate
instability via injection of contraction energy and diffusion of
free neutrals relative to freezed ions.

In this paper we conclude that linear thermal instability can
produce small-scale condensations in spherical, oblate, or
prolate. We try to analyze a rather involved problem linearly,
because, before this process overcome, \textit{turbulence} causes
to interaction and merging of these newly formed condensations.
Physically, we expect that merging of these small-scale
condensations culminate in the large-scale clumps that are star
bearing regions in our world. Authors are now preparing a
complete simulated turbulent magnetic molecular cloud with
condensations produced by thermal instability. It would be
interested to investigate the effect of interaction, merger, and
coagulation of these small-scale condensations with smoothed
particle hydrodynamics~(SPH) method.

\end{article}
\end{document}